# Energy-Efficient Micromixing in Paper Based Devices Mediated by the Interplay of Electrical and Thermal Fields

Golak Kunti, Sankha Shuvra Das, Vinay Manaswi Pedireddi, Anandaroop Bhattacharya, Suman Chakraborty[*]

Department of Mechanical Engineering, Indian Institute of Technology Kharagpur, Kharagpur, West Bengal - 721302, India

[*]*email: [suman@mech.iitkgp.ernet.in](suman@mech.iitkgp.ernet.in)*

Biomedical and biochemical processes in paper-based microfluidic devices often deal with mixing of two analytes to perform important functions. Uniform mixing of analytes in paper matrix is a challenging proposition, often necessitating complicated channel design or high energy external fields for realizing the desired functionality. In sharp contrast, here we demonstrate an energy-efficient technique compatible with handing biofluids, to achieve a high level of mixing of two fluids in paper-based microfluidic devices without deploying complex channel shapes. Our method employs a simple, cost-effective printing-based method to fabricate paper channel with interdigitated electrodes laid out using pencil sketch. An alternating current electric field of sufficiently low voltage is used to generate perturbations on the interface of the two fluids streams undertaking spontaneous capillary imbibition. The consequent variations in electric field generate local thermal gradients, leading to localized volumetric forces triggering efficient fluid mixing. Our results find a wide variety of applications ranging from biochemical analysis to medical diagnostics.

**INTRODUCTION**

In recent years, paper-based analytical devices (µPADs) have gained significant interest due to their advantageous features such as frugality, simple fabrication, portability, flexibility, ease in handling etc., with applications ranging from water or food quality monitoring,[1,2] energy generation and storage,[3–6] fluid mixing,[7] detection of biochemical entities,[8,9] disease diagnostics[10–12] etc. From a more fundamental perspective, these devices also provide an adequate surface to capture and visualize the analytes.[13] In addition, capillary pores in paper devices obviate the need of external pumping to drive the fluid, exploiting an intrinsic fluidic actuation.

Because of numerous advantages of µPADs, a number of fabrication techniques have been established in current years, including inkjet printing, photolithography, cutting, stamping, wax printing and screen-printing etc.[2,14–19,13–17] Recent fabrication techniques have also emphasized on means of altering the wettabilities of paper substrates.[20] These methods have been largely developed to achieve desired functionalities consistent with biochemical and biomedical tasks such as human blood typing, biomarker of pathogens test, plasma separation, immunochromatographic test etc.[21–24] In many of these applications, determination of color intensity of analytes predicts the detection of protein, glucose, uric acid, lactate, and cholesterol in µPADs.[15,25–28]

An important aspect of most of the biomedical and biochemical processes is that the functionality and effectiveness of the processes depend on the mixing of the analyte and the sample within the µPAD. Detection of bioreagent, ELISA-based pathogen detection etc. are thus associated with mixing of various analytes.[29,30] Naturally occurring diffusion-based passive mixing in the paper-fibre matrix, where analytes are transported by capillary action, has poor efficiency and poor controllability.[31,32] To achieve improved mixing through external field activation on paper platforms, various techniques have emerged so far, such as mixing using surface acoustic wave (SAW),[31] electric field[32] etc. Notably, SAW demands a high power budget (of the order of 1 W) to achieve the desired mixing performance.

Electric field actuated micromixing traditionally relies on electrical double layer (EDL) phenomenon triggering electrokinetic transport. Effectiveness of this process strongly depends on solution conductivity. Beyond a threshold conductivity ( > 0.085 S/m), electroosmotic velocity significantly drops due to compression of the EDL.[33] However, biological analytes typically possess high electrical conductivities (may go to few S/m). Accordingly, electroosmotic micromixers may not turn out to be effective for biological applications.[34] However, utilizing variations in electrical properties with temperature may be alternatively used as strategy for fluid pumping,[35–37] mixing,[34,38,39] two-phase flow [40–42] of high conductivity fluids.

Here, we show that by exploiting the electrical-field-gradient-driven interfacial perturbation of two fluid streams spontaneously evolved out of capillary imbibition, the resultant generation of non-uniform thermal field[43–45] may trigger sufficient driving mechanisms for highly efficient mixing on a paper matrix. Fundamentally, local gradients of the dielectric properties induce locally differential mobile charges, facilitating fluidic mixing mediated by differential electrical forcing. Unlike traditional electrokinetics, this technique is effective for high-conducting media and does not necessitate high electric field towards achieving the desired functionality. Moreover, simple channel layout can be used to achieve the desired task, unlike typical corrugated channel-pattern employed in several previously reported studies,[31,32] obviating the need of sophisticated fabrication strategies. Ideally suited for high-conductivity fluids, this method further appears to be appropriate for executing biological operations on a paper platform.[46–48] Further, being effective at sufficiently low voltages (<20 V), one may realize the paradigm of economised power budget.[49,50]

**EXPERIMENTAL SECTION**

**Device fabrication**

An ink-jet printing-based technique is used to fabricate the paper channel on a standard laboratory grade filter paper (Whatman Grade 4, mean pore diameter: 20-25 $\mu m$). This technique is much simpler and cost-effective compared to other printing-based technique, such as wax-printing. In this approach, designed channels are simply printed on both sides of the filter paper using an ink-jet printer. The printed paper channel is then heated on a hot-plate at 180-200 °C for 4-5 mins, which essentially melts the toner particles deposited at both the sides of the paper surface. The melted toner particles thus penetrate and block the paper pores, and hence create a hydrophobic barrier with distinct hydrophilic region. Thus, the printed ink based hydrophobic barrier provides directionality to the test liquid, which thereby imbibe through the hydrophilic section of the paper channel. Figure 1 illustrates the schematic and a fabricated straight paper channel (inset) with Y –inlet junction, used for present mixing study. To pattern the electrodes, HB graphite pencil is sketched following the electrode design. The electrodes are drawn from a distance of 5 mm away from the Y-junction of the channel (considering as an origin of the coordinate system i.e., $x = 0$). To illustrate the principle, we consider four pairs of parallel electrodes sketched along the width of the channel, with typical dimensions as follows: width ~ $628.5 \pm 19 \mu m$ and inter-electrode spacing ~ $243.8 \pm 14 \mu m$. The graphite electrode spans over a length of 10 mm. While we have performed experiments with several other combinations of these dimensions, we report only a limited set results here with an impression that the primary focus of the present study is not to bring out the resulting large set of parametric information but primarily to illustrate the essential principle on a paper platform and the nature of dependence on the design parameters. Copper wires are attached at both the electrode pads using conductive silver paste (Alfa Aesar: sheet resistance < 0.025 $\Omega$/square @ 0.001 in. thick). In order to apply an electric field, an AC function generator (33250A, Agilent) is connected with the copper electrodes, where AC signal of the function generator is amplified by 50 fold using an amplifier (High Voltage Wide Band Amplifier 9200).

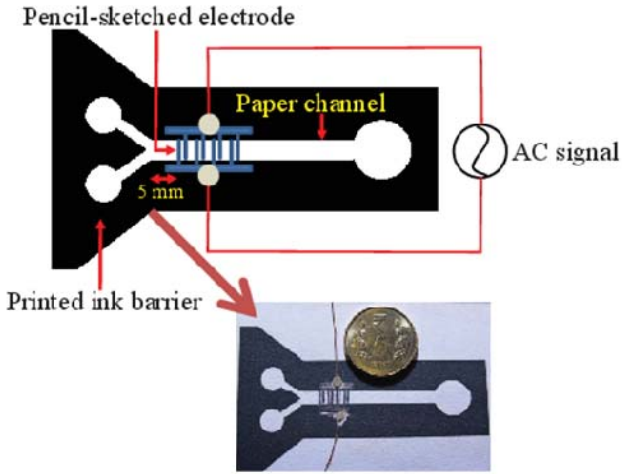

**Figure 1.** Schematic representation of the inkjet printed device. The barrier of the paper channel is made using the ink of the cartridge and HB pencil is used to draw the electrodes. Flow of the fluid is driven by natural wicking action. On activation of the electric field, perturbations in the fluids cause fluid mixing.

**Experimental methodology**

Mixing performance is characterized under various operating as well as geometrical parametric conditions. The primary factors influencing the mixing performance are the applied AC electric field and electrical conductivity of the fluid. In addition, mixing characteristics may also change with the alteration of geometrical parameters such as channel width. Here, we use KCl electrolyte solution as the background fluid media. Two different coloured dyes such as red and blue are mixed with electrolyte in the same volume ratio of 1:10 (v/v). During the experimentation, a constant volume of 20 µl coloured dyes is dispensed at both the inlet reservoirs. We consider two different cases, namely, with and without electric field, to evaluate the effect of the electric field on mixing performance. During the experimentation, images of the mixing process are captured using a digital camera (Nikon COOLPIX L810). All the experiments were performed several times to ensure the repeatability of the results and consistent statistical estimates.

**Characterization of mixing process**

Colour hue-based technique is adopted to quantify the mixing performance. Towards that, we have followed the methodology described in Ref.[31] This technique is well established nowadays and is adopted in majority of the detection analyses whose principles are based on colour change events.[51,52] In this technique, hue value of the pixel is normalized as:[31]

$$H^* = \frac{H - H_{mix}}{H_b - H_{mix}} \quad (1)$$

where $H^*$ and $H_b$ are the actual hue value and hue value of blue ink-stained electrolyte. $H_{mix}$ is the hue value of the equally mixed red- and blue-dyed solutions. Hue values of red-dyed solution and blue-dyed solution used for our experiments are are 360 and 240, respectively. To characterize the mixing performance, we have defined a mixing index as follow:

$$\gamma = 1 - H^* \quad (2)$$

For unmixed condition, one can obtain $H \approx H_b$ and $\gamma \approx 0$ (case of poor mixing). On the other hand, for complete mixing, $H \approx H_{mix}$ and $\gamma \approx 1$. An area of $10 \times 100$ pixels is taken for 4 mm width channel over which the mixing analysis is performed. Here, 100 pixels cover the channel width of 4 mm. Corresponding pixels for 2 mm and 6 mm width channels are 50 and 150, respectively. This area is taken at before the beginning of the electrode patterning ($x = 5\,\text{mm}$) and end of the electrode patterning ($x = 15\,\text{mm}$).

**NUMERICAL MODELING**

**governing transport equations**

The equation governing the electric potential (with negligible effect of magnetic field), i.e. the Laplace equation can be expressed as $\nabla^2 \varphi = 0$, $\mathbf{E} = -\nabla\varphi$, where $\varphi$ is the electric potential and $E$ is the electric field strength. The electrothermal body force can be expressed as: [43]

$$\mathbf{F_E} = \frac{1}{2}\varepsilon\,\mathrm{Re}\left[(\alpha-\beta)\frac{(\nabla T \cdot \mathbf{E})}{1+(\omega\tau)^2}\mathbf{E} - \frac{1}{2}\alpha\nabla T|\mathbf{E}|^2\right], \quad (3)$$

where $\omega = 2\pi f$ is the angular frequency of the AC signal, $\tau = \varepsilon/\sigma$ is charge relaxation time of the electrolyte and constants $\alpha$ and $\beta$ are $\frac{1}{\varepsilon}\frac{\partial \varepsilon}{\partial T} \approx -0.004\,°\mathrm{C}^{-1}$ and $\frac{1}{\sigma}\frac{\partial \sigma}{\partial T} \approx 0.02\,°\mathrm{C}^{-1}$,[53] respectively, $T$ is the temperature.

The temperature distribution due to electric field is governed by:

$$\rho C_p \mathbf{V}\cdot \nabla T = k\nabla^2 T + \sigma|\mathbf{E}|^2. \quad (4)$$

where $\rho, C_p, \mathbf{V},$ and $k$ are the density, specific heat, flow velocity and thermal conductivity of the fluid respectively, and $\sigma|\mathbf{E}|^2$ is the heat source caused by Joule heating.

We assume the flow to be steady, and incompressible. Therefore, the Navier-Stokes equation can be expressed as :

$$\rho(\mathbf{V}\cdot\nabla\mathbf{V}) - \mu\nabla^2\mathbf{V} + \nabla p = \mathbf{F_E}, \quad (5)$$

where $\mu$ is the dynamic viscosity of the fluid and $p$ is the pressure. The Navier-Stokes equation is solved in conjunction with mass conservation equation: $\nabla \cdot \mathbf{U} = 0$.
Diffusion and convection mechanisms govern the mass transport of a sample introduced into the flow field. At steady state, the sample concentration distribution can be expressed as:

$$\mathbf{U}\cdot\nabla C - D\nabla^2 C = 0, \quad (6)$$

where $C$ and $D$ are the concentration and mass diffusion coefficient of the sample.

**Numerical simulation**

We have used COMSOL Multiphysics to perform our numerical analysis. Symmetric pairs of electrodes have been placed on the bottom surface. The concept geometry is shown in Figure 2. In the figure, an expanded view of one symmetric pair of electrode is shown. In a representative model problem simulated with the sole objective of understanding the fluid flow details, the height and length of the channel have been taken as $H = 50$ μm and 10 mm, respectively. Four pairs of symmetric electrodes are connected on the bottom surface to stir the fluid. The gap between adjacent electrodes is 243 μm. Length of each electrode is 628 μm.

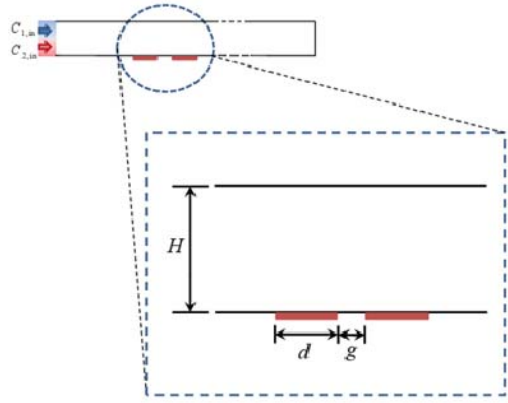

**Figure 2.** Schematic representation of one pair electrodes on the bottom wall for numerical simulation.

We have used KCl electrolyte to fill the microfluidic channel. Properties of the solution are: $\rho = 1000$ kg/m$^3$, $C_p = 4184$ J/kgK, $\mu = 0.00108$ Pa s, $k = 0.61$ W/mK, $\sigma = 0.56$ S/m, $\varepsilon = \varepsilon_o \varepsilon_r = 7.08 \times 10^{-10}$ C/Vm, where $\varepsilon_0$ is permittivity of vacuum and $\varepsilon_r$ is the relative permittivity of the liquid (i.e., $\varepsilon_0 = 8.85 \times 10^{-10}$ C/Vm, $\varepsilon_r = 80$), respectively. The diffusion coefficient for solute transport is taken as $10^{-10}$ m$^2$/s.[54]

We have applied an AC potential ($\pm V_{rms}$) at a frequency of $f$ and a phase difference of 180° between pairs of electrode. Other boundaries are electrically insulating. For the thermal field, the inlet is taken at the ambient condition and the exit condition is taken to be free of axial thermal gradient. Other boundaries are thermally insulated. For the velocity field, no slip and no penetration boundary conditions are applied on the solid walls. Fluid velocity at inlet is taken as $10^{-4}$ m/s and at outlet zero gauge pressure is set. We consider the sample concentrations at the inlets to be 1 mol/m$^3$ and 0 mol/m$^3$, whereas at the outlet of the channel, the condition of zero solute gradient along the axial direction is set.

## RESULTS AND DISCUSSIONS

### Characterization of mixing

Micromixing of biofluids in μPADs is based on the generation of disturbance against the capillary driven flow through the paper pores as mediated by the electrical field. In general, the applied electrical signal generates microvortices over a pair of symmetric electrodes (symmetric with respect to size, shape of the electrodes).[55] Symmetrically distributed potential lines and isothermal lines result in rotational flows over the electrodes. Volumetric heating due to the interplay of ionic current and electrical voltage occurs at the gap of the electrodes where electric field strength is high. Hence, vortices are located over the edges of the electrode near the electrode spacing. These vortices perturb the intrinsic capillary driven transport in paper channels. Uniform mixing can be achieved on the progression of the streams over sufficient electrode pairs.

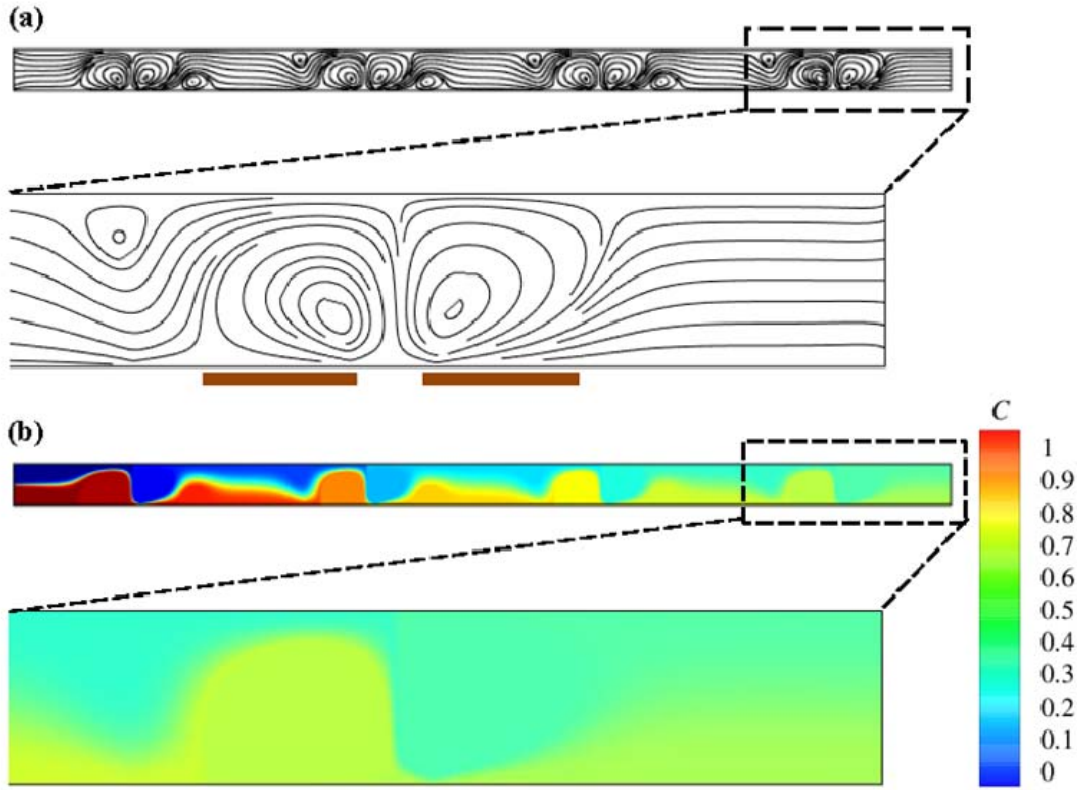

**Figure 3.** Numerical results showing (a) streamline distribution and (b) concentration distribution. Magnified view (dotted box) shows the detail of mixing. Two vortices are generated over one pair of electrodes that help to stir the fluids. Almost uniform mixing is achieved after crossing four electrode pairs.

Figure 3 shows the numerical results of distributions of streamlines and concentration of samples (concentration varies from 0 to 1 mol/m$^3$ across the channel width) to understand the flow dynamics. From Eq. (3) we can state that fluid motion is altered at locations where temperature gradient prevails. The maximum temperature occurs at inter-electrode gap, and decreases away from that location. Forces due to electrical perturbation on the fluid stream, mediated by temperature-dependence of electrical properties, generate two vortices above the gap of the electrodes (Figure 3a). The enlarged view of the dotted box containing one pair of electrode is shown and positions of the electrodes are also shown separately. Strong vortices significantly perturb the smoothly progressed flow patterns of the two streams and stir the liquids. The contact area of the liquid streams is enhanced above the electrode pairs due to the rotating motion of the fluids. This results in almost uniform mixing at the outlet of the channel (Figure 3b).

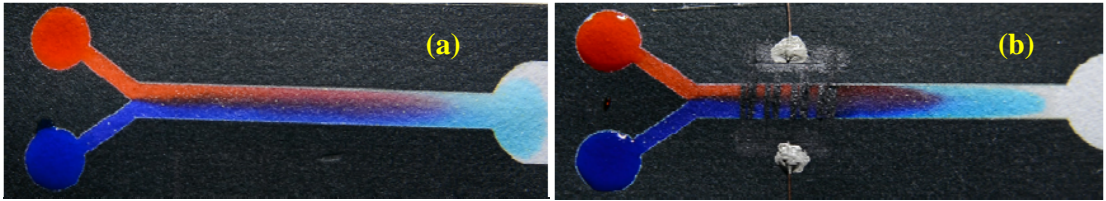

**Figure 4.** Captured experimental images showing mixing performance for two different cases: (a) without electric field and (b) using an AC electric field (voltage 13.25 $V_{rms}$, conductivity 1.92 S/m). The channel width is 4 mm for both the reported cases. Mixing index shows an appreciable enhancement in mixing performance upon application of electric field.

To identify the significant effects of electrical perturbation driven mechanism on the mixing process, we show captured images for two different cases: mixing through natural capillary imbibition and mixing using electrothermal forces (refer, Figure 4). Images were captured after a time interval of 600 s. It is important to mention that we started measuring time as both the fluids arrived at the first electrode. The applied voltage and conductivity values are ~13.25 V (RMS value) and ~1.92 S/m, respectively, for electrically mediated mixing. Mixing analysis is carried out at the region located after the electrode patterning ($x = 15\,\text{mm}$). The mixing index obtained for two different cases (electrically aided and passive) are $0.64 \pm 0.019$ and $0.28 \pm 0.013$, respectively. Therefore, a significant enhancement in mixing quality is achieved using electrical forcing.

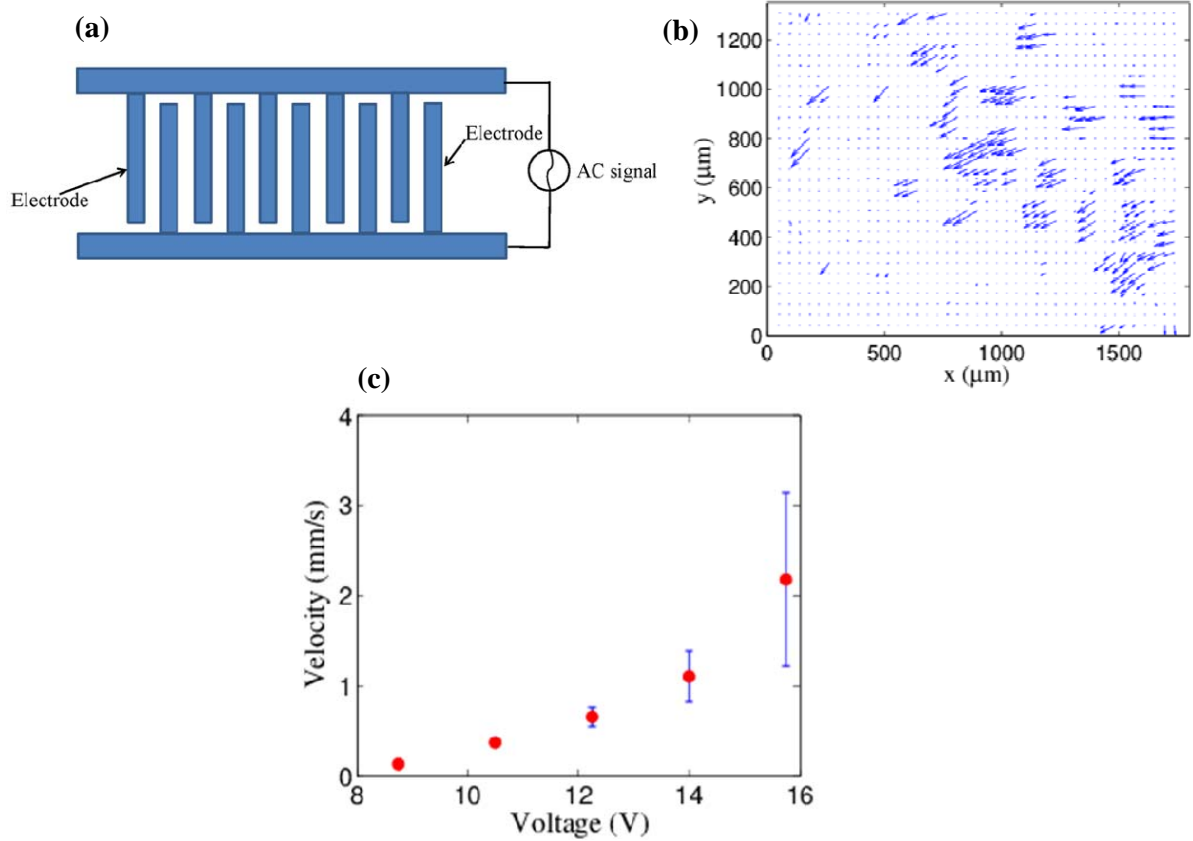

**Figure 5.** (a) Symmetric electrode arrangement to generate the flow, (b) Velocity vectors of the flow field at applied voltage of 14 $V_{rms}$ and conductivity of 1.92 S/m. Symmetric electrode patterning imposes a curvilinear motion on the fluid, which perturbs the two streams of fluids flowing over the electrodes, (c) variation of average velocity with voltages. Standard deviation of the velocity data is shown by the *error bars*. Fluid velocity rapidly increases with applied voltage and shows power law dependence of $U \sim \varphi^{4.88}$ with applied voltage.

**Effect of voltage**

The experimentally obtained flow over symmetrical electrodes shows a curvilinear motion over the electrode surface. Images were taken using the fluorescence microscopy (Olympus IX71) on a paper substrate. The velocity field over the pencil sketched electrode (Figure 5a) is adopted to characterize the flow dynamics. The vector plot and average fluid velocity as a function of voltage are highlighted in Figure 5b and 5c, respectively. The curvilinear fluid motion over the electrode surface is effective to alter the flow direction when fluid flows over the paper substrate. The fluid velocity rapidly increases with applied voltage. Balance between viscous force and electrical force yields flow velocities scaled as $U \sim \nabla T \times E^2$. Further, considering conduction-dominated thermal transport, $\nabla T \sim E^2$. Thus, coupled effects of non-uniform thermal and electric fields cause a fluid velocity which has theoretically fourth power dependence with applied voltage ($U \sim E^4, U \sim \varphi^4$, since $E \sim \varphi/g$, $\varphi$ and $g$ are the voltage and gap between the electrode, respectively). However, our experimentally obtained index of velocity dependence on voltage is 4.88, which slightly overshoots the theoretical estimation. This deviation can be attributed to the fact that during the experiment, evaporation of the electrolyte reduces the fluid volume. However, salt in the solution remains the same and concentration increases. As a result, the volumetric heating gradually increases in reduced volume of fluid with time and fluid velocity increases.

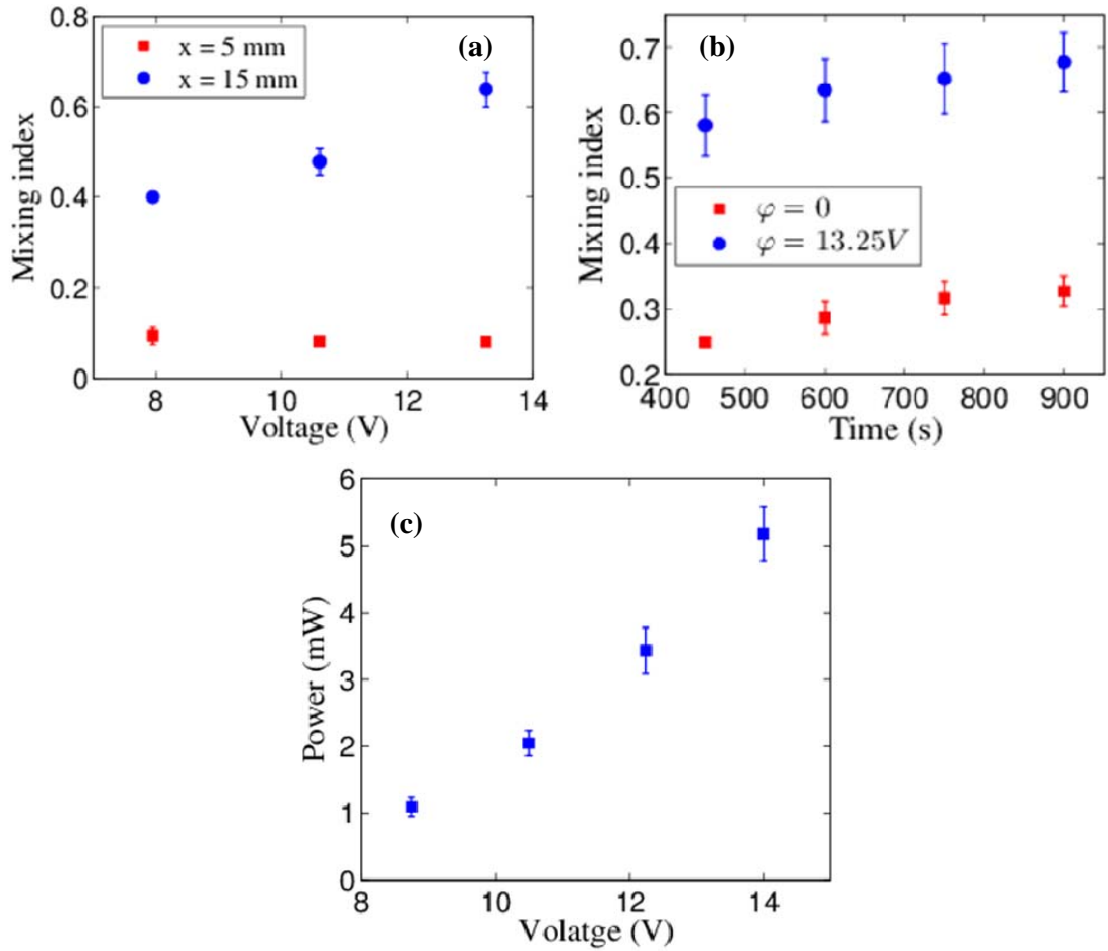

**Figure 6.** (a) experimental data of mixing index for various applied voltages ($\varphi$ = 7.95-13.25 V$_{rms}$) at locations ($x$ = 5 and 15 mm). A case of without applied voltage is also considered to characterize the mixing performance. Mixing index increases significantly with increased voltage (b) mixing index versus time plot for two cases: mixing with only capillary action and mixing using combined effect of capillarity and electrical force (c) power budget estimation for different voltages. Electrical power required to drive electrically mediated mixing in paper-based microfluidic system is of the order of milli Watt. Electrical conductivity and channel width for (a)-(c) are 1.92 S/m and 4 mm, respectively.

This effect is neglected in the theoretical and scaling estimation. In our experiments, voltage of 7.95-14 V was used to characterize the mixing process. Below $\varphi_{rms} = 7.95\,\text{V}$, significant mixing was not observed. Voltage above 14 V for $\sigma = 1.92$ S/m was not considered in our experiments, due to the formation of bubbles which occurs owing to electrolysis reaction above a threshold voltage.[56] This threshold voltage can be increased at lower conductivities. For $\sigma = 1.03$ and 0.56, threshold voltages were found 21.21V and 26.52V, respectively. All the experiments were performed at a frequency of 500 kHz, exploring a regime where temperature-dependent volumetric forces dominate significantly over the other electrokinetic forces[57] Moreover, the possibility of bubble formation is obviated at such frequencies.[56] Figure 6a shows the variation of mixing index for various voltages for $\sigma = 1.92$ S/m at time $t = 600\,s$. Values of mixing index are shown at two different locations such as before the electrode patterning ($x = 5\,\text{mm}$) and after the electrode patterning ($x = 15\,\text{mm}$). We also consider a control case of zero applied voltage. Under such circumstances, mixing is completely facilitated by capillary wicking. Non-uniform distribution of paper-fibre forms a matrix of non-uniform pores. Fluid flow due to wicking action takes place randomly, enabling passive mixing of the two dyed solutions. However, quality of the mixed sample is not good. One can notice mixing index below 0.3 is obtained without electric field (Figure 5b). Nevertheless, once electric field is applied, local gradients of dielectric properties due to volumetric heating induce forces on the fluid located in the paper pores. This results in local stretching of fluid elements, thereby increasing interfacial area of contact between two fluids. The rapidity of change of flow directions through the paper-fibre increases with increasing voltage and mixing index shows increasing trend with voltage. Mixing index reaches a value of 0.64 at a voltage of 13.25 V (Figure 6a). Because of absence of flow perturbations, the mixing index is very low ($< 0.1$) before electrode patterning. The mixing index also depends on the time elapsed of the dyed solutions during the passage of electrode (see Figure6b). Electrical forces increase the area of contact of the streams which further increases with time. Although the mixing of fluids through internal diffusion and externally applied disturbance increases with time, the rate of enhancement of mixing index using the electrically mediated scheme is higher as compared to that through diffusion process.

We also measure the electrical power required to run the micromixer. Data of electrical power with varied voltage is presented in Figure 6c. Estimated power for mixing the dyed streams is of the order of milli Watt, which is three order of magnitude less as compared to the previously reported mixing processes using surface acoustic wave (SAW) on paper matrix.

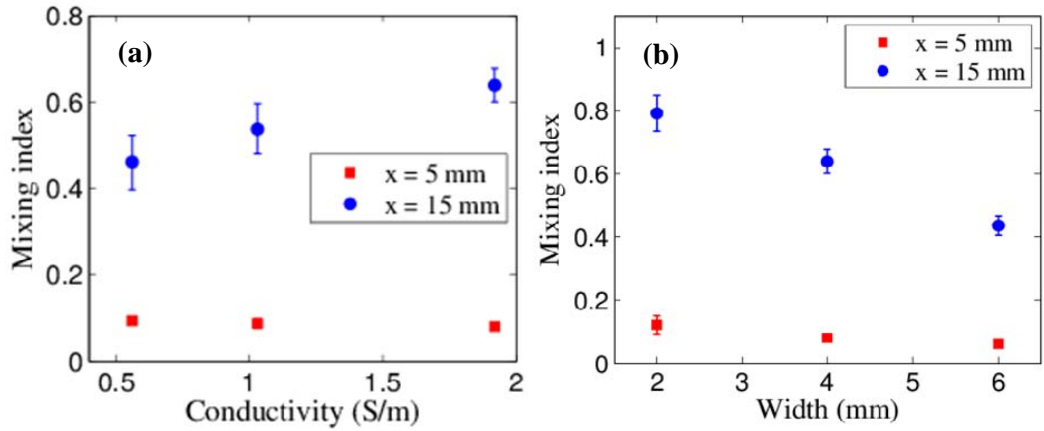

**Figure 7.** Variation in mixing index at different locations ($x = 5, 15$ mm) for different (a) electrical conductivities ($\sigma = 0.56 - 1.92$ S/m) and channel widths ($w = 2 - 6$ mm). Applied voltage and channel width for the cases reported in (a) are: $13.25\,V_{rms}$ and 4 mm respectively, while the applied voltage and electrical conductivity for the cases reported in (b) are $13.25\,V_{rms}$ and 1.92 S/m, respectively. Data of mixing index are calculated at $t = 600$ s.

**Effect of electrical conductivity and channel width**

Figure 7a shows the trend of variation of mixing index with electrical conductivity of the fluid at different locations of the channel, for an applied voltage of 13.25 V. The mixing index shows increasing trend with electrical conductivity, which is favorable for biological and biochemical processes. Enhancement in mixing index with electrical conductivity can be attributed to the fact that volumetric heating, which generates the thermal field, increases with electrical conductivity. This, in turn, amplifies the local flow perturbations, facilitating fluidic mixing. It is further seen that the mixing efficiency increases at reduced channel width (Figure 7b). The contact area of the streams cannot travel throughout the entire width of the channel for larger width. Virtually uniform mixing can be achieved for channel width of 2 mm (mixing index 0.79). Therefore, smaller channel width is preferable to obtain a uniform mixing. Mixing index for channel width 6 mm shows poor mixing at the end of the mixing process.

**CONCLUSIONS**

A new strategy of fabricating paper-based micromixer is presented by deploying the thermal field generated out of electrically-modulated interfacial perturbations in capillary-driven fluid steams. The fabricated device is simple to fabricate, cost-effective and easily disposable, facilitating the handling of biological samples. Furthermore, applicability to handle fluids of high conductivity and enhanced mixing performance with increases in the same ensures the appropriateness for manipulation of biofluids. The channel width can be further kept to a minimum within the fabrication constraints to achieve the best outcome. These results may open up new possibilities of energy-efficient point-of-care diagnostics in resource-limited settings.

**ACKNOWLEDGMENTS**

This research has been supported by Indian Institute of Technology Kharagpur, India [Sanction Letter no.: IIT/SRIC/ATDC/CEM/2013-14/118, dated 19.12.2013].